\documentclass[proceedings]{JHEP3}
\usepackage{amsfonts}
\usepackage{amsmath}
\usepackage{epsfig}

\newbox\mybox

\newcommand\fverb{\setbox\mybox=\hbox\bgroup\verb}
\newcommand\fverbdo{\egroup\medskip\noindent\fbox{\unhbox\mybox}\ }
\newcommand\fverbit{\egroup\item[\fbox{\unhbox\mybox}]}
\conference{q-deformed states for generalized uncertainty relations}
\abstract{We investigate properties of generalized time-dependent q-deformed coherent states for a
noncommutative harmonic oscillator. The states are shown to satisfy a generalized version 
of Heisenberg's uncertainty relations. For the initial value in time the states are
demonstrated to be squeezed, i.e. the inequalities are saturated, whereas when time evolves 
the uncertainty product oscillates away from this value albeit still respecting the relations. For the canonical variables
on a noncommutative space we verify explicitly that Ehrenfest's theorem hold at all times. We conjecture that the model exhibits 
revival times to infinite order. Explicit sample computations for the fractional revival times and superrevival times are presented.}

\title{Time-dependent q-deformed coherent states for generalized uncertainty
relations}
\author{Sanjib Dey$^\bullet$, Andreas Fring$^\bullet$, Laure Gouba$^\circ$
and Paulo G. Castro$^\bullet$ \\
$^\bullet$ Department of Mathematical Science, City University London,\\
$\,\,$ Northampton Square, London EC1V 0HB, UK\\
$^\circ$ The Abdus Salam International Centre for Theoretical Physics
(ICTP), \\
$\,\,$ Strada Costiera 11, 34014, Trieste, Italy \\
E-mail: sanjib.dey.1@city.ac.uk, a.fring@city.ac.uk, lgouba@ictp.it,
paulo.castro.1@city.ac.uk}

\input{tcilatex}
\begin{document}

\section{Introduction}

The algebras satisfied by the canonical variables resulting from $q$%
-deformed oscillator algebras have been shown to be related to
noncommutative spacetime structures leading to minimal lengths and minimal
momenta as a result of a generalized version of Heisenberg's uncertainty
relations \cite{AFBB,AFLGFGS,AFLGBB,DFG}. An important question to address
in this context is whether explicit states satisfying these relations
actually exist and how they can be constructed. Recently two of the present
authors \cite{DF2} have investigated this problem for a nontrivial limit of
the $q$-deformed oscillator algebra. Using generalized coherent states,
so-called Klauder-coherent states \cite{Klauder0,Klauder01,Klauder1,Klauder2}%
, it was shown in \cite{DF2} for a noncommutative harmonic oscillator to
first order perturbation theory in the deformation parameter that these
states not only satisfy the generalized uncertainty relations, but even
saturate them at all times. The main purpose of this paper is to extend this
type of analysis to the case for generic deformation parameter $q$.

\section{Generalized time-dependent q-deformed coherent states}

Following \cite{Arik,Bieden,MacF,ChangPu,Kulish:1990eh}, up to minor
differences in the conventions, we consider a one dimensional $q$-deformed
oscillator algebra for the creation and annihilation operators $A^{\dagger }$
and $A$ in the form 
\begin{equation}
AA^{\dagger }-q^{2}A^{\dagger }A=1,\quad \ \ \ \ \ \text{for }q\leq 1.
\label{AAAA}
\end{equation}%
Defining a $q$-deformed version of the Fock space involving q-deformed
integers $[n]_{q}$ as%
\begin{equation}
\left\vert n\right\rangle _{q}:=\frac{\left( A^{\dagger }\right) ^{n}}{\sqrt{%
[n]_{q}!}}\left\vert 0\right\rangle ,\quad \lbrack n]_{q}:=\frac{1-q^{2n}}{%
1-q^{2}},\quad \lbrack n]_{q}!:=\dprod\limits_{k=1}^{n}[k]_{q},\quad
A\left\vert 0\right\rangle =0,\quad \left\langle 0\right\vert \!\left.
0\right\rangle =1,  \label{Fock}
\end{equation}%
it follows immediately that the operators $A^{\dagger }$ and $A$ act indeed
as raising and lowering operators, respectively,%
\begin{equation}
A^{\dagger }\left\vert n\right\rangle _{q}=\sqrt{[n+1]_{q}}\left\vert
n+1\right\rangle _{q},\quad \quad \text{and\quad \quad }A\left\vert
n\right\rangle _{q}=\sqrt{[n]_{q}}\left\vert n-1\right\rangle _{q}.
\end{equation}%
Furthermore, one deduces from (\ref{AAAA}) and (\ref{Fock}) that the states $%
\left\vert n\right\rangle _{q}$ form an orthonormal basis, i.e. $%
_{q}\!\left\langle n\right\vert \!\left. m\right\rangle _{q}=\delta _{n,m}$.
As was first argued in \cite{Arik}, the $q$-deformed Hilbert space $\mathcal{%
H}_{q}$ is then spanned by the vectors $\left\vert \psi \right\rangle
:=\dsum\nolimits_{n=0}^{\infty }c_{n}\left\vert n\right\rangle _{q}$ with $%
c_{n}\in \mathbb{C}$, such that $\left\langle \psi \right\vert \!\left. \psi
\right\rangle =\dsum\nolimits_{n=0}^{\infty }\left\vert c_{n}\right\vert
^{2}<\infty $.

Using these states we can construct the Klauder-coherent states introduced
in \cite{Klauder0,Klauder01,Klauder1,Klauder2}. In general, these states are
defined for a Hermitian Hamiltonian $H$ with discrete bounded below and
nondegenerate eigenspectrum and orthonormal eigenstates $\left\vert \phi
_{n}\right\rangle $ as a two parameter set 
\begin{equation}
\left\vert J,\gamma \right\rangle =\frac{1}{\mathcal{N}(J)}%
\sum\limits_{n=0}^{\infty }\frac{J^{n/2}\exp (-i\gamma e_{n})}{\sqrt{\rho
_{n}}}\left\vert \phi _{n}\right\rangle ,\text{\qquad }J\in \mathbb{R}%
_{0}^{+},\gamma \in \mathbb{R}.  \label{GK}
\end{equation}%
The probability distribution and normalization constant 
\begin{equation}
\rho _{n}:=\dprod\limits_{k=1}^{n}e_{k},\qquad \text{and\qquad }\mathcal{N}%
^{2}(J):=\dsum\limits_{k=0}^{\infty }\frac{J^{k}}{\rho _{k}},  \label{N2}
\end{equation}%
are expressed in terms of the scaled energy eigenvalues $e_{n}$ resulting
from $H\left\vert \phi _{n}\right\rangle =\hbar \omega e_{n}\left\vert \phi
_{n}\right\rangle $. The key properties of these states are their continuity
in the two variables $(J,\gamma )$, the fact that they provide a resolution
of the identity and that they are temporarily stable satisfying the action
angle identity $\left\langle J,\gamma \right\vert H\left\vert J,\gamma
\right\rangle =\hbar \omega J$. The time evolution is governed by a shift in
the parameter $\gamma $, i.e. $\exp (-iHt/\hbar )\left\vert J,\gamma
\right\rangle =\left\vert J,\gamma +t\omega \right\rangle $.

As a concrete system let us now consider the noncommutative harmonic
oscillator Hamiltonian $H=\hbar \omega A^{\dagger }A$, where the operators $%
A^{\dagger }$ and $A$ obey (\ref{AAAA}). With the re-scaled eigenvalues $%
e_{n}=[n]_{q}$ and eigenstates $\left\vert \phi _{n}\right\rangle
=\left\vert n\right\rangle _{q}$ for this Hamiltonian, we obtain the
probability distribution $\rho _{n}=[n]_{q}!$. Furthermore, the
normalization condition $\left\langle J,\gamma \right. \!\!\!$ $\left\vert
J,\gamma \right\rangle =1$ yields the $q$-deformed exponential $E_{q}(J)$ as
the normalization constant 
\begin{equation}
E_{q}(J):=\dsum\limits_{n=0}^{\infty }\frac{J^{n}}{[n]_{q}!}=\mathcal{N}%
^{2}(J).
\end{equation}%
Thus our normalized coherent state 
\begin{equation}
\left\vert J,\gamma \right\rangle _{q}:=\frac{1}{\sqrt{E_{q}(J)}}%
\sum\limits_{n=0}^{\infty }\frac{J^{n/2}\exp (-i\gamma e_{n})}{\sqrt{[n]_{q}!%
}}\left\vert n\right\rangle _{q},  \label{coho}
\end{equation}%
coincides with the coherent state $\left\vert z\right\rangle $, as defined
already in \cite{Arik}, for the specific choice $\left\vert
z^{2},0\right\rangle _{q}$, that is for $t=0$. Let us now investigate some
properties of these states and in particular investigate to which kind of
expectation values they lead for observables and compare with the results
for the nontrivial $q\rightarrow 1$ limit studied in \cite{DF2}. In the
latter case these states were found to be squeezed states up to first order
in perturbation theory in $\tau $ when parameterizing the deformation
parameter as $q^{\tau }$. Most importantly we wish to investigate whether
these states respect the generalized uncertainty relations.

\section{Generalized Heisenberg's uncertainty relations}

In order to verify the uncertainty relations projected onto these states we
commence by recalling \cite{AFBB,Kempf2,DFG} that the canonical variables
expressed in terms of the $q$-deformed oscillator algebra generators%
\begin{equation}
X=\alpha \left( A^{\dagger }+A\right) ,\qquad \text{and\qquad }P=i\beta
\left( A^{\dagger }-A\right) ,  \label{XP}
\end{equation}%
with $\alpha =1/2\sqrt{1+q^{2}}\sqrt{\hbar /(m\omega )}$ and $\beta =1/2%
\sqrt{1+q^{2}}\sqrt{\hbar m\omega }$, satisfy the deformed canonical
commutation relations%
\begin{equation}
\lbrack X,P]=i\hbar +i\frac{q^{2}-1}{q^{2}+1}\left( m\omega X^{2}+\frac{1}{%
m\omega }P^{2}\right) .  \label{one}
\end{equation}%
The interesting feature about this version of a noncommutative spacetime is
that it leads to a minimal length as well as a minimal momentum. Let us
first analyze the generalized version of Heisenberg's uncertainty relation
for a simultaneous measurement of the two observables $X$ and $P$ projected
onto the normalized coherent states $\left\vert J,\gamma \right\rangle _{q}$
as defined in equation (\ref{coho}) 
\begin{equation}
\left. \Delta X\Delta P\right\vert _{\left\vert J,\gamma \right\rangle
_{q}}\geq \frac{1}{2}\left\vert \left( _{q}\!\left\langle J,\gamma
\right\vert [X,P]\left\vert J,\gamma \right\rangle _{q}\right) _{\eta
}\right\vert .  \label{GHU}
\end{equation}%
The uncertainty for $X$ is computed as $\Delta X^{2}=\left(
_{q}\!\left\langle J,\gamma \right\vert X^{2}\left\vert J,\gamma
\right\rangle _{q}\right) _{\eta }-\left( _{q}\!\left\langle J,\gamma
\right\vert X\left\vert J,\gamma \right\rangle _{q}\right) _{\eta }^{2}$ and
analogously for $P$ with $X\rightarrow P$. The $\eta $ indicates that we
might have to change to a nontrivial metric when $X$ and/or $P$ are
non-Hermitian following the prescriptions provided in the recent literature
on non-Hermitian systems \cite{Urubu,Bender:1998ke,AliI,Benderrev,Alirev} or
more specifically for this particular setting in \cite{DF2}.

Notice that when we assume that the conjugation of $A$ and $A^{\dagger }$
yield $A^{\dagger }$ and $A$, respectively, the operators $X$ and $P$ can be
seen as Hermitian. In that case the metric $\eta $ is taken to be the
standard one, possibly with some change to ensure proper self-adjointness
and the convergence of the inner products. Indeed, in \cite{MacF,Atak} such
a representation on a unit circle acting on Rogers-Sz\"{e}go polynomials 
\cite{Andrews} was derived 
\begin{equation}
A=\frac{i}{\sqrt{1-q^{2}}}\left( e^{-i\check{x}}-e^{-i\check{x}/2}e^{2\tau 
\check{p}}\right) ,\qquad \text{and\qquad }A^{\dagger }=\frac{-i}{\sqrt{%
1-q^{2}}}\left( e^{i\check{x}}-e^{2\tau \check{p}}e^{i\check{x}/2}\right) .
\label{rep1}
\end{equation}%
Here we used the dimensionless parameters $\check{x}=x\sqrt{m\omega /\hbar 
\text{ }}$ and $\check{p}=p/\sqrt{m\omega \hbar \text{ }}$ with $x$, $p$
being the standard canonical coordinates satisfying $\left[ x,p\right]
=i\hbar $. Evidently $A^{\dagger }$ is the conjugate of $A$ for $q<1$ and
consequently with (\ref{XP}) follows that also the canonical variables
satisfying (\ref{one}) are Hermitian in this representation, i.e. $%
X^{\dagger }=X$, $P^{\dagger }=P$. We notice further that for the
representation (\ref{rep1}) the $\mathcal{PT}$-symmetry of the standard
canonical variables $\mathcal{PT}$: $x\rightarrow -x$, $p\rightarrow p$, $%
i\rightarrow -i$ is inherited by canonical variables on the noncommutative
space $\mathcal{PT}$: $X\rightarrow -X$, $P\rightarrow P$, $i\rightarrow -i$.

There exist also alternative representations \cite{Burban} 
\begin{equation}
A=\frac{1}{1-q^{2}}D_{q},\qquad \text{and\qquad }A^{\dagger
}=(1-x)-x(1-q^{2})D_{q},  \label{Dq}
\end{equation}%
in terms of Jackson derivatives $D_{q}f(x):=[f(x)-f(q^{2}x)]/[x(1-q^{2})]$
introduced in \cite{Jack}. The operators in (\ref{Dq}) commute to (\ref{AAAA}%
) when acting on eigenvectors constructed from normalized Rogers-Sz\"{e}go
polynomials. It is less obvious to see whether this representation can be
made Hermitian. For our purposes it is important that at least one such
representation exists and we may compute expectation values on the $q$%
-deformed Fock space with the standard metric.\qquad

In order to verify the inequality (\ref{GHU}) for the states (\ref{coho}) we
compute first the expectation values for the creation and annihilation
operators%
\begin{equation}
_{q}\!\left\langle J,\gamma \right\vert A\left\vert J,\gamma \right\rangle
_{q}=J^{1/2}\frac{F_{q}(J,-\gamma )}{E_{q}(J)},\qquad \text{and\qquad }%
_{q}\!\left\langle J,\gamma \right\vert A^{\dagger }\left\vert J,\gamma
\right\rangle _{q}=J^{1/2}\frac{F_{q}(J,\gamma )}{E_{q}(J)},  \label{AA}
\end{equation}%
where we introduced the function 
\begin{equation}
F_{q}(J,\gamma ):=\dsum\limits_{n=0}^{\infty }\frac{J^{n}e^{i\gamma q^{2n}}}{%
[n]_{q}!}=\dsum\limits_{n=0}^{\infty }\frac{i^{n}}{n!}E_{q}(q^{2n}J)\gamma
^{n}.
\end{equation}%
Notice that this function reduces to the $q$-deformed exponential $%
F_{q}(J,0)=E_{q}(J)$ and also the duality in the derivatives with respect to
the two parameters. The standard derivative with respect to $\gamma $
corresponds to a $q$-deformation in the parameter $J$%
\begin{equation}
-i\frac{d}{d\gamma }F_{q}(J,\gamma )=F_{q}(q^{2}J,\gamma )  \label{ID1}
\end{equation}%
and in turn the Jackson derivative acting on $J$ is identical to a
deformation in the second parameter%
\begin{equation}
D_{q}F_{q}(J,\gamma )=\frac{F_{q}(J,\gamma )-F_{q}(q^{2}J,\gamma )}{%
J(1-q^{2})}=F_{q}(J,q^{2}\gamma ).  \label{ID2}
\end{equation}%
These identities are easily derived from the defining relations for $F_{q}$
and will be made use of below. Using the representations for $X$ and $P$ in
terms of the creation and annihilation operators (\ref{XP}), it follows
directly with the help of (\ref{AA}) that%
\begin{eqnarray}
_{q}\!\left\langle J,\gamma \right\vert X\left\vert J,\gamma \right\rangle
_{q} &=&\frac{\alpha J^{1/2}}{E_{q}(J)}\left[ F_{q}(J,\gamma
)+F_{q}(J,-\gamma )\right] , \\
_{q}\!\left\langle J,\gamma \right\vert P\left\vert J,\gamma \right\rangle
_{q} &=&\frac{i\beta J^{1/2}}{E_{q}(J)}\left[ F_{q}(J,\gamma
)-F_{q}(J,-\gamma )\right] .
\end{eqnarray}%
To compute the expectation values for $X^{2}$ and $P^{2}$, we use once again
(\ref{XP}) to express them in terms of the $A^{\dagger }$ and $A$. Thus we
evaluate 
\begin{eqnarray}
_{q}\!\left\langle J,\gamma \right\vert A^{\dagger }A^{\dagger }\left\vert
J,\gamma \right\rangle _{q} &=&J\frac{F_{q}(J,\gamma (1+q^{2}))}{E_{q}(J)},
\label{AA1} \\
_{q}\!\left\langle J,\gamma \right\vert AA\left\vert J,\gamma \right\rangle
_{q} &=&J\frac{F_{q}(J,-\gamma (1+q^{2}))}{E_{q}(J)}, \\
_{q}\!\left\langle J,\gamma \right\vert A^{\dagger }A\left\vert J,\gamma
\right\rangle _{q} &=&J, \\
_{q}\!\left\langle J,\gamma \right\vert AA^{\dagger }\left\vert J,\gamma
\right\rangle _{q} &=&1+q^{2}J,  \label{AA4}
\end{eqnarray}%
and with $X^{2}=\alpha ^{2}(A^{\dagger }A^{\dagger }+A^{\dagger
}A+AA^{\dagger }+AA)$ and $P^{2}=-\beta ^{2}(A^{\dagger }A^{\dagger
}-A^{\dagger }A-AA^{\dagger }+AA)$ we assemble this to 
\begin{eqnarray}
_{q}\!\left\langle J,\gamma \right\vert X^{2}\left\vert J,\gamma
\right\rangle _{q} &=&\alpha ^{2}\left[ J\frac{F_{q}(J,\gamma
(1+q^{2}))+F_{q}(J,-\gamma (1+q^{2}))}{E_{q}(J)}+1+J+q^{2}J\right] , \\
_{q}\!\left\langle J,\gamma \right\vert P^{2}\left\vert J,\gamma
\right\rangle _{q} &=&-\beta ^{2}\left[ J\frac{F_{q}(J,\gamma
(1+q^{2}))+F_{q}(J,-\gamma (1+q^{2}))}{E_{q}(J)}-1-J-q^{2}J\right] .~~~
\end{eqnarray}%
From these expressions we find that the right hand side of the generalized
Heisenberg's inequality (\ref{GHU}) is always a constant value independent
of $\gamma $, i.e. time, 
\begin{equation}
\frac{1}{2}\left\vert _{q}\!\left\langle J,\gamma \right\vert \hbar +\frac{%
q^{2}-1}{q^{2}+1}\left( m\omega X^{2}+\frac{1}{m\omega }P^{2}\right)
\left\vert J,\gamma \right\rangle _{q}\right\vert =\frac{\hbar }{4}(1+q^{2})%
\left[ 1+(q^{2}-1)J\right] .  \label{RHS}
\end{equation}%
The square of the left hand side of (\ref{GHU}) \ can be written as 
\begin{equation}
\left. \Delta X^{2}\Delta P^{2}\right\vert _{\left\vert J,0\right\rangle
_{q}}=\alpha ^{2}\beta ^{2}\left[ 1+(1+q^{2})J+G_{q}-G_{c}^{2}(\gamma )%
\right] \left[ 1+(1+q^{2})J-G_{q}-G_{s}^{2}(\gamma )\right] ,  \label{LHS}
\end{equation}%
where we introduced the functions%
\begin{equation}
G_{c}(\gamma ):=\frac{2\sqrt{J}}{E_{q}(J)}\dsum\limits_{n=0}^{\infty }\frac{%
J^{n}}{[n]_{q}!}\cos (\gamma q^{2n}),~\qquad ~\text{~}G_{s}(\gamma ):=\frac{%
2i\sqrt{J}}{E_{q}(J)}\dsum\limits_{n=0}^{\infty }\frac{J^{n}}{[n]_{q}!}\sin
(\gamma q^{2n}),  \label{G}
\end{equation}%
and $G_{q}:=\sqrt{J}G_{c}(\gamma +\gamma q^{2})$. Noting that $%
\lim\nolimits_{\gamma \rightarrow 0}G_{q}=2J$, $\lim\nolimits_{\gamma
\rightarrow 0}G_{c}(\gamma )=2\sqrt{J}$ and $\lim\nolimits_{\gamma
\rightarrow 0}G_{s}(\gamma )=0$, it is easy to see that for $\gamma =0$ the
expression (\ref{LHS}) becomes the square of (\ref{RHS}), such that the
minimal uncertainty product for the observables $X$ and $P$ is saturated.
From the expressions in (\ref{G}) we deduce that the range for these
functions is $-2J\leq G_{q}\leq 2J$, $0\leq G_{c}^{2}(\gamma )\leq 4J$ and $%
-4J\leq G_{s}^{2}(\gamma )\leq 0$. Recognizing next that the inequality
holds when each of the brackets in (\ref{LHS}) is greater than $1+(q^{2}-1)J$%
, this requires that $2J\geq G_{c}^{2}(\gamma )-G_{q}$ and at the same time $%
2J\geq G_{s}^{2}(\gamma )+G_{q}$. This means $4J\geq G_{c}^{2}(\gamma
)+G_{s}^{2}(\gamma )$, which by the previous estimates is indeed the case.
Overall this implies that for $\gamma \neq 0$ the uncertainty relation (\ref%
{GHU}) is always respected.

Next we verify Ehrenfest's theorem. For the time evolution of the operator $%
X $ we compute directly%
\begin{equation}
i\hbar \frac{d}{dt}~_{q}\!\left\langle J,\omega t\right\vert X\left\vert
J,\omega t\right\rangle _{q}=-\frac{\omega \hbar \alpha J^{1/2}}{E_{q}(J)}%
\left[ F_{q}(q^{2}J,\omega t)-F_{q}(q^{2}J,-\omega t)\right] ,  \label{E1}
\end{equation}%
and compare it to%
\begin{equation}
_{q}\!\left\langle J,\omega t\right\vert \left[ X,H\right] \left\vert
J,\omega t\right\rangle _{q}=-\frac{\omega \hbar \alpha J^{1/2}}{E_{q}(J)}%
\dsum\limits_{s=\pm \omega t}\frac{s}{\omega t}F_{q}(J,s)+\frac{s}{\omega t}%
J(q^{2}-1)F_{q}(J,q^{2}s),  \label{E2}
\end{equation}%
which is easily computed from the expectation values%
\begin{eqnarray}
_{q}\!\left\langle J,\gamma \right\vert A^{\dagger }A^{\dagger }A\left\vert
J,\gamma \right\rangle _{q} &=&J^{3/2}\frac{F_{q}(J,q^{2}\gamma )}{E_{q}(J)},
\label{AAA1} \\
_{q}\!\left\langle J,\gamma \right\vert A^{\dagger }AA^{\dagger }\left\vert
J,\gamma \right\rangle _{q} &=&J^{1/2}\frac{F_{q}(J,\gamma )}{E_{q}(J)}%
+q^{2}J^{3/2}\frac{F_{q}(J,q^{2}\gamma )}{E_{q}(J)}, \\
_{q}\!\left\langle J,\gamma \right\vert A^{\dagger }AA\left\vert J,\gamma
\right\rangle _{q} &=&J^{3/2}\frac{F_{q}(J,-q^{2}\gamma )}{E_{q}(J)}, \\
_{q}\!\left\langle J,\gamma \right\vert AA^{\dagger }A\left\vert J,\gamma
\right\rangle _{q} &=&J^{1/2}\frac{F_{q}(J,-\gamma )}{E_{q}(J)}+q^{2}J^{3/2}%
\frac{F_{q}(J,-q^{2}\gamma )}{E_{q}(J)}\text{.}  \label{AAA4}
\end{eqnarray}%
The equality of (\ref{E1}) and (\ref{E2}) follows from the identities (\ref%
{ID1}) and (\ref{ID2}). Similarly we verified the validity of Ehrenfest's
theorem also for the operator $P$.

\section{Revival times}

As previously argued \cite{Perel,Klauder2,DF2}, revival time structures are
very interesting and important quantities of time dependent states as in
principle they are measurable quantities, see for instance \cite{Mol}. The
structure is directly linked to the dependence of the energy eigenvalues $%
E_{n}$ on the quantum number $n$, i.e. the existence of the $k$-th
derivative $d^{k}E_{\bar{n}}/d\bar{n}^{k}$ with respect to some average
value $\bar{n}$ at which the wave packet $\psi =\dsum c_{n}\phi _{n}$ is
well localized. For the case at hand these derivatives exist to all orders,
such that we expect infinitely many revival times to exist.

\begin{figure}[h]
\centering   \includegraphics[width=4.9cm]{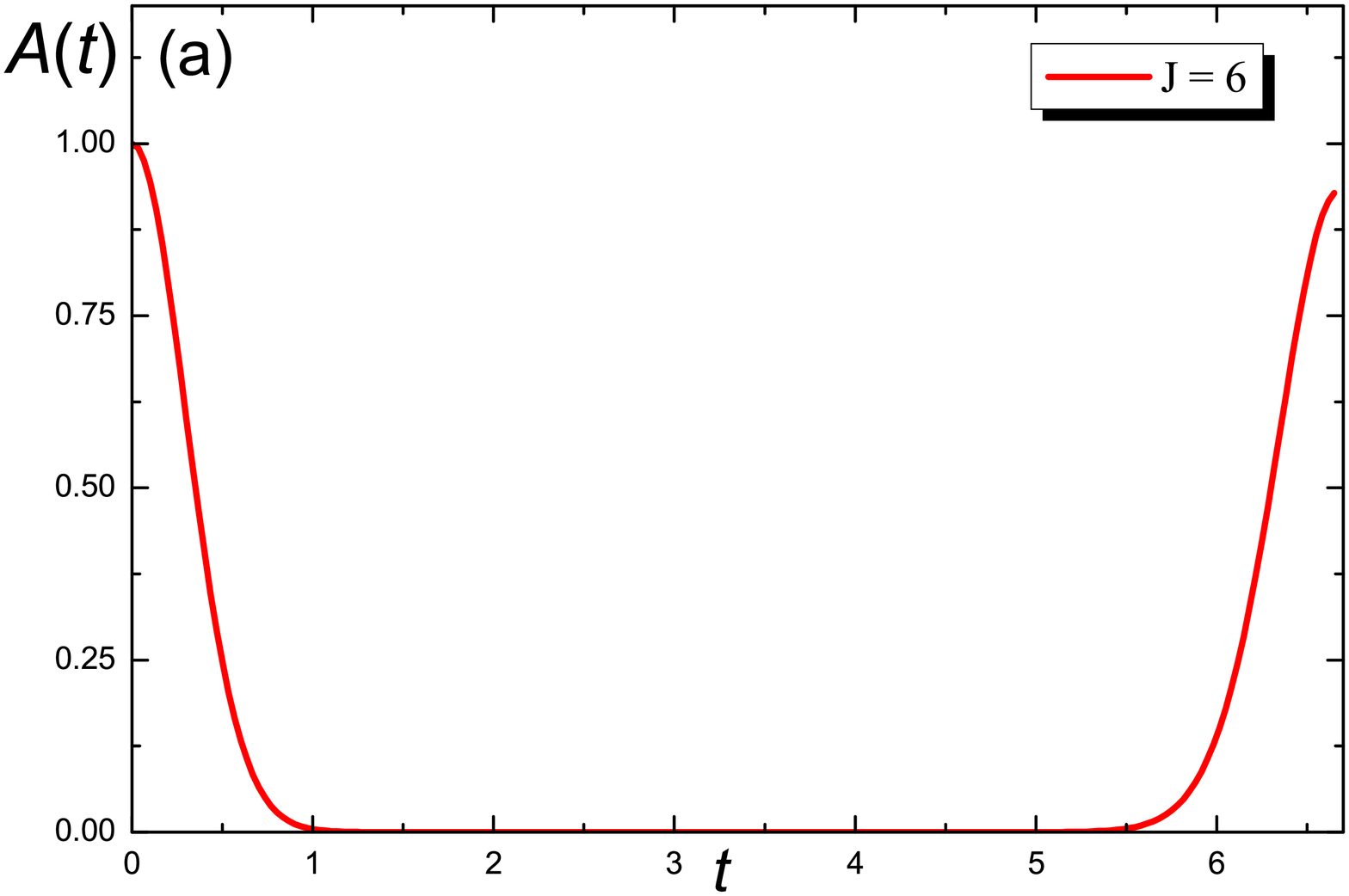} %
\includegraphics[width=4.9cm]{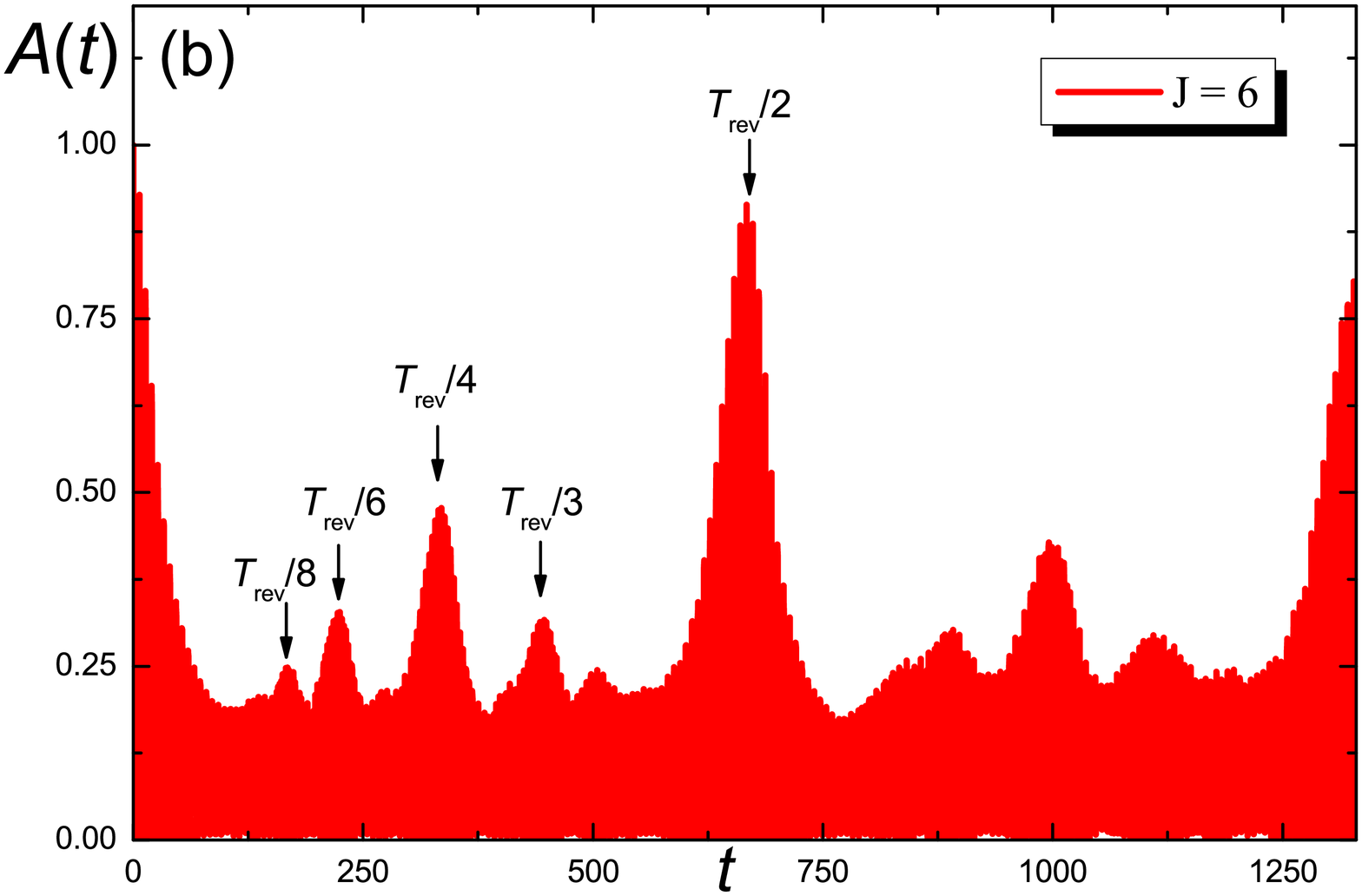} %
\includegraphics[width=4.9cm]{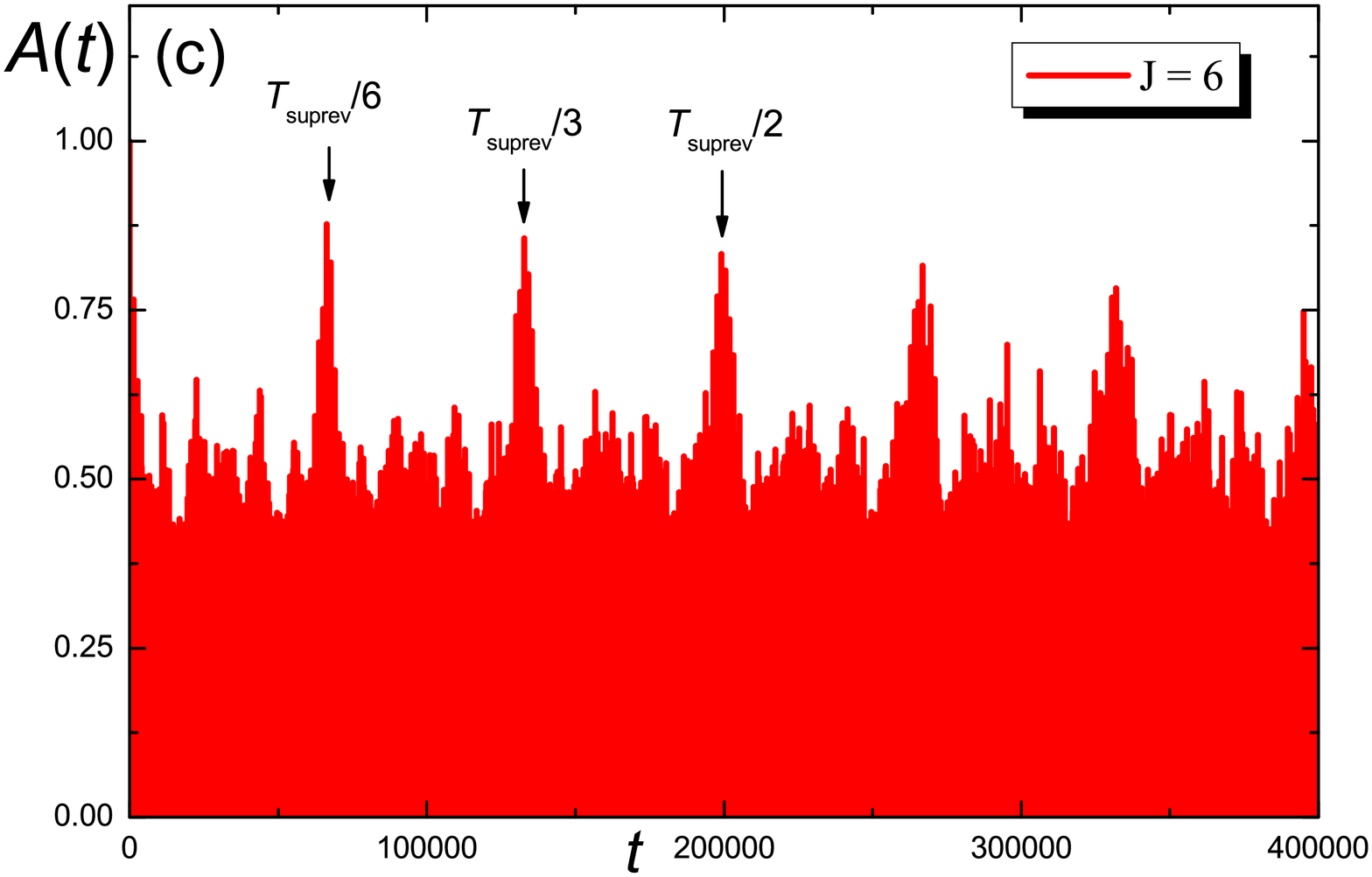}
\caption{Autocorrelation function as a function of time at different scales
for $\hbar=1$, $\protect\omega= 1$, $q=e^{-0.005}$, $J=6$ and $\bar{n}%
=6.1875 $. (a) Classical period at $T_{\text{cl}}=6.65$, (b) fractional
revival times for $T_{\text{rev}}=1330.19$ and (c) fractional superrevival
times for $T_{\text{suprev}}=3999056$. }
\label{F1}
\end{figure}

At the smallest scale one obtains the classical period $T_{\text{cl}}=2\pi
\hbar /\left\vert E_{\bar{n}}^{\prime }\right\vert $, thereafter at larger
scale the fractional revivals for the revival time $T_{\text{rev}}=4\pi
\hbar /\left\vert E_{\bar{n}}^{\prime \prime }\right\vert $, then the
superrevival time $T_{\text{suprev}}=12\pi \hbar /\left\vert E_{\bar{n}%
}^{\prime \prime \prime }\right\vert $, etc. For the case at hand the peak
of the wave packet is computed to $\bar{n}:=\left\langle n\right\rangle
=Jd\ln \mathcal{N}^{2}(J)/dJ$. Noting that $d^{k}E_{n}/dn^{k}=\hbar \omega
2^{k}q^{2n}\ln ^{k}q/(q^{2}-1)$ we obtain the times 
\begin{equation}
T_{\text{cl}}=\frac{\pi }{\omega }\left\vert \frac{q^{2}-1}{q^{2\bar{n}}\ln q%
}\right\vert ,\quad \quad T_{\text{rev}}=\frac{\pi }{\omega }\left\vert 
\frac{q^{2}-1}{q^{2\bar{n}}\ln ^{2}q}\right\vert ,\qquad \text{and\quad
\quad }T_{\text{suprev}}=\frac{3\pi }{2\omega }\left\vert \frac{q^{2}-1}{q^{2%
\bar{n}}\ln ^{3}q}\right\vert .  \label{TT}
\end{equation}%
In figure 1 we present the autocorrelation function as a function of time at
different scales. In panel (a) the revival after the classical period is
clearly visible. The parameters have been chosen in a way that $T_{\text{rev}%
}/T_{\text{cl}}\approx 200$, such that at the revival time scale the
revivals due to the classical periods have died out and only the revival due
to $T_{\text{rev}}$ are exhibited as clearly visible in the computation
presented in panel (b). With $T_{\text{suprev}}/T_{\text{rev}}\approx 300$
this type of behaviour is repeated at the superrevival time scale as seen in
panel (c). Due to the aforementioned dependence of the energy eigenvalues on 
$n$, we conjecture here that this behaviour is repeated order by order.
However, the verification of this feature poses a more and more challenging
numerical problem which we leave for future investigations.

\section{Conclusions}

By extending the analysis of \cite{DF2}, from a perturbative treatment to
the generic case for $q<1$, we have computed time dependent $q$-deformed
coherent states for a harmonic oscillator on a noncommutative space. We
demonstrated that all key requirements for coherent states are satisfied. A
direct comparison with the results obtained in \cite{DF2} is not possible as
the analysis in there relates to a nontrivial limit $q\rightarrow 1$, which
is not directly obtainable from the setting presented here. However,
qualitatively we found a somewhat different behaviour with regard to the key
question addressed in this manuscript. Whereas the perturbative treatment in 
\cite{DF2} indicated a saturation for the generalized version of
Heisenberg's uncertainty relation at all times, the generic $q$-deformed
states exhibit this feature only for $t=0$, but do respect the inequality
thereafter. We have also presented explicit computations for the
verification of Ehrenfest's theorem for the coordinate and momentum operator
at all times. By computing the autocorrelation functions we have shown that
besides a fractional revival time structure this system also exhibits a
superrevival structure at a much larger time scale.

Clearly there are various open problems left for future investigations, such
as the study of different types of models on the type of noncommutative
spaces investigated here. Especially an extension to higher dimensional
models would be very interesting. It would also be interesting to study
representations for which the operators $X$ and $P$ are non-Hermitian, as
for instance in (\ref{Dq}), in analogy to the analysis presented in \cite%
{DF2}. More computational power should also allow to confirm our conjecture
about the existence of revival time structure at much larger time scales,
such as supersuperrevival time structures, etc.

\bigskip \noindent \textbf{Acknowledgments:} SD is supported by a City
University Research Fellowship. LG is supported by the high energy section
of the ICTP. PGC is supported by CNPq.


\end{document}